\documentclass[preprint,showkeys]{revtex4}

\usepackage{graphics}
\usepackage{graphicx}
\usepackage{amssymb}
\usepackage{amsmath}
\usepackage{amsfonts}
\usepackage{setspace}

\newif\ifFIG
\FIGtrue

\begin{document}   

\author{Hiroo Nozaki}
\author{Yuji Ikeda}
\author{Kazuhide Ichikawa}
\author{Akitomo Tachibana}
\email{akitomo@scl.kyoto-u.ac.jp}
\affiliation{Department of Micro Engineering, Kyoto University, Kyoto, 615-8540, Japan}

\title{Electronic stress tensor analysis of molecules in gas phase of CVD process for GeSbTe alloy}


\begin{abstract}
We analyze the electronic structure of molecules which may exist in gas phase of 
chemical vapor deposition process for GeSbTe alloy using the electronic stress tensor,
with special focus on the chemical bonds between Ge, Sb and Te atoms.
We find that, from the viewpoint of the electronic stress tensor, they have intermediate
properties between alkali metals and hydrocarbon molecules. 
We also study the correlation between the bond order which is defined based on the electronic stress tensor,
and energy-related quantities. 
We find that the correlation with the bond dissociation energy is not so strong while one with
the force constant is very strong. We interpret these results in terms of the energy density
on the ``Lagrange surface", which is considered to define the boundary surface of atoms in a molecule
in the framework of the electronic stress tensor analysis. 
\end{abstract}


\maketitle

\section{Introduction} \label{sec:intro}

GeSbTe (GST) alloy is the most popular material for phase change memory (PCM) \cite{Terao2009,Burr2010},
which is one of the most promising candidates for the next-generation memory device. 
So far, GST thin films are deposited by physical vapor deposition such as sputtering and pulsed laser deposition. 
However, chemical vapor deposition (CVD) of GST \cite{Kim2006,Kim2009,Machida2010} has many advantages such 
as good step coverage, uniformity and high purity,
and considered to be necessary for future PCM applications. 
The CVD process for GST is relatively a new field of research and there remains many things which are not clearly known. 
One of them is the chemical reactions in the gas phase of CVD process, and we have investigated reactions and molecules which 
may exist in the gas phase using quantum chemical calculation \cite{PCOS2010}.
As the result of this study, we have obtained a data set of molecules which have bonds among Ge, Sb and Te.
In this paper, we apply our electronic stress tensor analysis based on the Rigged QED (Quantum Electrodynamics) theory \cite{Tachibana2001,Tachibana2003,Tachibana2004,Tachibana2005,Tachibana2010,Tachibana2013,Tachibana2014} 
to this data set, and investigate how these bonds can be described by the electronic stress tensor.

In general, the stress tensor, which describes a pattern of internal forces of matter, is widely used in  
various fields of science such as mechanical engineering and material science. 
The use of stress tensor in quantum systems as well has been investigated for many years, 
by one of the earliest quantum mechanics papers \cite{Schrodinger1927, Pauli} and many researchers
\cite{Epstein1975, Bader1980, Bamzai1981a, Nielsen1983, Nielsen1985, Folland1986a, Folland1986b, Godfrey1988, Filippetti2000, Pendas2002, Rogers2002, Morante2006, Tao2008, Ayers2009, Jenkins2011,GuevaraGarcia2011}
including our group \cite{Tachibana2001,Tachibana2003,Tachibana2004,Tachibana2005, Szarek2007, Szarek2008, Szarek2009, Ichikawa2009a, Ichikawa2009b,Tachibana2010, Ichikawa2010, Ichikawa2011a, Ichikawa2011b, Ichikawa2011c, Henry2011,Ichikawa2012,Tachibana2013,Tachibana2014}. 
In our past studies, we have shown that the electronic stress tensor and related quantities can be useful tools to analyze atomic and molecular systems and can give new viewpoints on the nature of chemical bonding. 

We here introduce two of our findings which we wish to investigate more deeply using GST bonds in this paper. 
First, we have pointed out that 
we may characterize some aspects of a metallic bond or metallicity of a chemical bond 
in terms of the electronic stress tensor \cite{Szarek2007,Ichikawa2012}.
We have analyzed the bonding region in the small cluster models and periodic models of Li and Na using the electronic stress tensor,
and have found that all the three eigenvalues of the stress tensor are negative and degenerate, just like those of liquid.
This is in stark contrast to hydrocarbon molecules, which have the positive largest eigenvalue much larger than
the other two negative eigenvalues.
The eigenvalue pattern of Li and Na indicates a lack of directionality and compressive nature of bonding while
that of hydrocarbon molecules indicates solid directionality and tensile nature of bonding.
Each pattern well reflects the nature of metallic and covalent bonding. 
Then, the question worth asking is, how the chemical bonds between 
metalloid atoms like Ge, Sb and Te are described by the electronic stress tensor.
Second, using the energy density which is defined from the electronic stress tensor,
new definition of bond order has been proposed \cite{Szarek2007}.
So far, we have investigated a correlation between this new bond order and bond distance \cite{Szarek2007,Szarek2008, Szarek2009,Ichikawa2011c}, and 
have found that our bond order exhibits better correlation than the other bond orders proposed in the literature. 
We, however, have not investigated a correlation with a quantity related to energy.
Therefore, we wish to investigate a correlation between our bond order
and the bond dissociation energy, which is available in our data set as the dimerization energy of the CVD precursors.
In addition, we compute a force constant for the GST bonds and examine a correlation with our bond order. 

The structure of this paper is as follows. 
Sec.~\ref{sec:theory} summarizes our analysis method of electronic structures using the electronic stress tensor density.
In Sec.~\ref{sec:results}, we show the electronic stress tensor density and its eigenvalues of the GST bonds,
and compare with those of hydrocarbon molecules and alkali metal clusters. 
We also investigate correlations between our bond order and energy-related quantities. 
Finally, Sec.~\ref{sec:conclusion} is devoted to our conclusion.

\section{Theory} \label{sec:theory}

In the following section, we analyze chemical bonds using quantities such as the electronic stress tensor density and the kinetic energy density. 
They are based on the Rigged QED theory \cite{Tachibana2003} and we briefly describe them in this section. 
The Rigged QED is a quantum field theory which has been proposed \cite{Tachibana2003} to treat dynamics of charged particles and photons in atomic and molecular systems. In addition to the ordinary QED which contains the Dirac field for electrons and the gauge field for photons, the Schr\"{o}dinger fields for atomic nuclei are included. 
More details are found in our previous papers \cite{Tachibana2001,Tachibana2003,Tachibana2010,Tachibana2013,Tachibana2014}.
Below, $c$ denotes the speed of light in vacuum, $\hbar$ the reduced Planck constant, 
$e$ the electron charge magnitude (so that $e$ is positive), and $m$ the electron mass. 
The gamma matrices are denoted by $\gamma^\mu$ ($\mu=$0-3).

The most basic quantity in the Rigged QED is the electronic stress tensor density operator $\hat{\tau}^{\Pi\, kl}_e(x)$  which is defined as follows \cite{Tachibana2001}.
\begin{eqnarray}
\hat{\tau}^{\Pi\, kl}_e(x) = \frac{i\hbar c}{2} 
\left[
\hat{\bar{\psi}}(x) \gamma^l \hat{D}_{e\, k}(x) \hat{\psi}(x) 
- \left( \hat{D}_{e\, k}(x) \hat{\psi}(x)\right)^\dagger \gamma^0 \gamma^l \hat{\psi}(x)
\right],   \label{eq:op_stress}
\end{eqnarray}
where $\hat{\psi}(x)$ is the four-component Dirac field operator for electrons, the dagger as a superscript is used to express Hermite conjugate,
and $\hat{\bar{\psi}}(x) \equiv \hat{\psi}^\dagger(x) \gamma^0$.
 We denote the spacetime coordinate as $x=(ct, \vec{r})$. The Latin letter indices like $k$ and $l$ express space coordinates from 1 to 3.
Here, the gauge covariant derivative is defined by
$\hat{D}_{e\, k}(x) = \partial_k + i\frac{Z_e e}{\hbar c} \hat{A}_k(x)$, where $Z_e = -1$,
and 
$\hat{A}_k(x)$ is the vector potential of the photon field operator in the Coulomb gauge (${\rm div} \hat{\vec{A}}(x)=0$). The important property of this quantity is that the time derivative of the electronic kinetic momentum density operator $\hat{\vec{\Pi}}_e(x)$
\begin{eqnarray}
\hat{\vec{\Pi}}_e(x) = \frac{1}{2} \left(
i\hbar \hat{\psi}^\dagger(x) \hat{\vec{D}}_e(x) \hat{\psi}(x) - i\hbar \left( \hat{\vec{D}}_e(x)\hat{\psi}(x) \right)^\dagger \cdot \hat{\psi}(x) \right),
\end{eqnarray}
can be expressed by the sum of the Lorentz force density operator $\hat{\vec{L}}_e(x)$ and the tension density operator $\hat{\vec{\tau}}^\Pi_e(x)$, which is the divergence of the stress tensor density operator:
\begin{eqnarray}
\frac{\partial}{\partial t} \hat{\vec{\Pi}}_e(x) = \hat{\vec{L}}_e(x) + \hat{\vec{\tau}}^\Pi_e(x).   \label{eq:dPiedt}
\end{eqnarray}
These operators are expressed as follows,
\begin{eqnarray}
\hat{\vec{L}}_e(x) &=& \hat{\vec{E}}(x) \hat{\rho}_e(x)  + \frac{1}{c} \hat{\vec{j}}_e(x) \times \hat{\vec{B}}(x),  \\
\hat{\tau}^{\Pi k}_e(x) &=& \partial_l \hat{\tau}^{\Pi\, kl}_e(x) \\
&=&
\frac{i\hbar c}{2} \Bigg[ \left(\hat{D}_{el}(x)\hat{\psi}(x) \right)^\dagger \gamma^0 \gamma^l \cdot \hat{D}_{ek}(x) \hat{\psi}(x)
+ \hat{\bar{\psi}}(x) \gamma^l \hat{D}_{ek}(x) \hat{D}_{el}(x) \hat{\psi}(x)  \nonumber \\
& &-\left( \hat{D}_{ek}(x)\hat{D}_{el}(x)\hat{\psi}(x) \right)^\dagger \gamma^0 \gamma^l \cdot \hat{\psi}(x)
-\left( \hat{D}_{ek}(x)\hat{\psi}(x) \right)^\dagger \gamma^0 \gamma^l \cdot \hat{D}_{el}(x) \hat{\psi}(x) \Bigg] \label{eq:op_tension} \nonumber \\
& &
-\frac{1}{c} \left( \hat{\vec{j}}_e(x) \times \vec{B}(x) \right)^k,
\end{eqnarray}
where $\hat{\vec{E}}(x)$ and $\hat{\vec{B}}(x)$ denote the electric field operator and magnetic field operator respectively, 
and $\hat{\rho}_e(x)$ and $\hat{\vec{j}}_e(x)$ are the electronic charge density operator and charge current density operator respectively. 

For nonrelativistic systems, we approximate the expressions above in the framework of
the Primary Rigged QED approximation \cite{Tachibana2013,Tachibana2014}, in which the small components of the four-component electron field are expressed by the large components as $\hat{\psi}_S (x) \approx  - \frac{1}{2m c} i\hbar \sigma^k  D_k \hat{\psi}_L (x)$ and the spin-dependent terms are ignored. Then, we take the expectation value of Eq.~\eqref{eq:dPiedt} with respect to the stationary state of the electrostatic Hamiltonian. This leads to the equilibrium equation as
\begin{eqnarray}
0 = \langle \hat{\vec{L}}_e(x) \rangle + \langle \hat{\vec{\tau}}^S_e(x) \rangle= \langle \hat{\vec{L}}_e(x) \rangle + \partial_l \langle  \hat{\tau}^{S\, kl}_e(x) \rangle ,	\label{eq:L}
\end{eqnarray}
which shows the balance between electromagnetic force and quantum field force at each point in space. Since this expresses the fact that the latter force keeps the electrons in the stationary bound state in atomic and molecular systems, we can expect to study the nature of chemical bonding from the viewpoint of quantum field theory by using the stress tensor density and tension density. We express $\langle \hat{\vec{\tau}}_e^S(x) \rangle$ and $\langle \hat{\vec{\tau}}_e^{Skl}(x) \rangle$ respectively $\tau^{Sk}(\vec{r})$ and $\tau^{Skl}(\vec{r})$ for simplicity (we also write only spatial coordinate $\vec{r}$ because we consider stationary state). The explicit expression for the stress tensor density $\tau^{Skl}(\vec{r})$  and tension density $\tau^{S k}(\vec{r})$ are
\begin{eqnarray} 
\tau^{Skl}_{e}(\vec{r}) &=& \frac{\hbar^2}{4m}\sum_i \nu_i
\Bigg[\psi^*_i(\vec{r})\frac{\partial^2\psi_i(\vec{r})}{\partial x^k \partial x^l}-\frac{\partial\psi^*_i(\vec{r})}{\partial x^k} \frac{\partial\psi_i(\vec{r})}{\partial x^l} \nonumber\\
& & \hspace{4cm} +\frac{\partial^2 \psi^*_i(\vec{r})}{\partial x^k \partial x^l}\psi_i(\vec{r}) -\frac{\partial \psi^*_i(\vec{r})}{\partial x^l}\frac{\partial \psi_i(\vec{r})}{\partial x^k}\Bigg], \label{eq:stress}
\end{eqnarray}
\begin{eqnarray} 
\tau^{S k}_{e}(\vec{r})&=&  \sum_l \partial_l  \tau^{Skl}(\vec{r}) \nonumber \\
&=&\frac{\hbar^2}{4m}\sum_i \nu_i
\Bigg[\psi^*_i(\vec{r})\frac{\partial \Delta\psi_i(\vec{r})}{\partial x^k}-\frac{\partial\psi^*_i(\vec{r})}{\partial x^k} \Delta\psi_i(\vec{r}) \nonumber\\
& & \hspace{4cm} +\frac{\partial \Delta\psi^*_i(\vec{r})}{\partial x^k}\psi_i(\vec{r}) -\Delta \psi^*_i(\vec{r}) \frac{\partial \psi_i(\vec{r})}{\partial x^k}\Bigg],
\label{eq:tension}
\end{eqnarray}
where $\psi_i(\vec{r})$ is the $i$th natural orbital and $\nu_i$ is its occupation number.
$\Delta$ denotes the Laplacian, $\Delta \equiv \sum_{k=1}^3 (\partial/\partial x^k)^2$.
When the density functional theory (DFT) method is used to compute the electronic structure, 
we use the Kohn-Sham orbitals for $\psi_i(\vec{r})$ in the above expressions.
The eigenvalue of the symmetric tensor $\stackrel{\leftrightarrow}{\tau}^{S}$ is the principal stress and the eigenvector is the principal axis  as follows:
\begin{eqnarray}
\stackrel{\leftrightarrow}{\tau}^{S}_{e}(\vec{r}) &=&
\begin{pmatrix}
\tau^{S}_{e\, xx}(\vec{r})&\tau^{S}_{e\, xy}(\vec{r})&\tau^{S}_{e\, xz}(\vec{r})\\
\tau^{S}_{e\, yx}(\vec{r})&\tau^{S}_{e\, yy}(\vec{r})&\tau^{S}_{e\, yz}(\vec{r})\\
\tau^{S}_{e\, zx}(\vec{r})&\tau^{S}_{e\, zy}(\vec{r})&\tau^{S}_{e\, zz}(\vec{r})
\end{pmatrix} \\
&\xrightarrow{{\rm diag}}&
\begin{pmatrix}
\tau^{S 11}_{e}(\vec{r})&0&0\\
0&\tau^{S 22}_{e}(\vec{r})&0\\
0&0&\tau^{S 33}_{e}(\vec{r})
\end{pmatrix},
\tau^{S11}_{e}(\vec{r}) \le \tau^{S22}_{e}(\vec{r}) \le \tau^{S33}_{e}(\vec{r}).
\label{eq:tensorge}
\end{eqnarray}

We use a concept of ``Lagrange point" \cite{Szarek2007} to characterize a bond between two atoms.
The Lagrange point $\vec{r}_L$ is defined as the point where the tension density $\vec{\tau}^S(\vec{r})$ vanishes,
namely $\tau^{S k}(\vec{r}_L)=0$.
We analyze chemical bonds by computing the eigenvalues of electronic stress tensor density at this point.

Another important quantity in the Rigged QED is the electronic kinetic energy density operator defined as
\cite{Tachibana2001},
\begin{eqnarray}
\hat{T}_e(x) &=& -\frac{\hbar^2}{2m}\cdot \frac{1}{2} \left( \hat{\psi}^\dagger(x) \hat{\vec{D}}^2_e(x) \hat{\psi}(x)
+ \left( \hat{\vec{D}}^2_e(x) \hat{\psi}(x) \right)^\dagger \cdot \hat{\psi}(x) \right). \label{eq:op_ked}
\end{eqnarray}
As is done for the electronic stress tensor density operator, 
we apply the Primary Rigged QED approximation to Eq.~\eqref{eq:op_ked} and take the expectation value with respect to the stationary state of the electrostatic Hamiltonian.
Then, we obtain the definition for the electronic kinetic energy density as
\begin{eqnarray}
n_{T_{e}}(\vec{r}) = - \frac{\hbar^2}{4m}
\sum_{i} \nu_i \left[ \psi_{i}^{*}(\vec{r}) \Delta \psi_{i}(\vec{r}) + 
\Delta \psi_{i}^{*}(\vec{r}) \cdot \psi_{i}(\vec{r}) \right].   \label{eq:ked} 
\end{eqnarray}
Note that our definition of the electronic kinetic energy density is not positive-definite. Using this kinetic energy density, we can divide the whole space into three types of region: the electronic drop region $R_D$ with $n_{T_{e}}(\vec{r}) > 0$, where classically allowed motion of electron is guaranteed and the electron density is amply accumulated; the electronic atmosphere region $R_A$ with $n_{T_{e}}(\vec{r}) < 0$, where the motion of electron is classically forbidden and the electron density is dried up; and the electronic interface $S$ with $n_{T_{e}}(\vec{r}) = 0$, the boundary between $R_D$ and $R_A$, which corresponds to a turning point. The $S$ can give a clear image of the intrinsic shape of atoms and molecules and is, therefore, an important region in particular. 

Finally, in our analysis, we use the energy density $\varepsilon_\tau^S(\vec{r})$ 
which is defined as a half of the trace of the stress tensor density \cite{Tachibana2001}:
\begin{eqnarray}
\varepsilon_\tau^S(\vec{r}) = \frac{1}{2} \sum_{k=1}^3 \tau^{Skk}(\vec{r}). \label{eq:ed}
\end{eqnarray}
It can be regarded as the energy density in a sense that the integration over whole space gives usual total energy $E$ of the system: $\int \varepsilon_\tau^S(\vec{r}) d\vec{r} = E$. This can be proved by using the virial theorem .
Using this energy density $\varepsilon_\tau^S(\vec{r})$, a new definition of the bond order
(bond strength index) is proposed \cite{Szarek2007}.
It is defined as $\varepsilon_\tau^S(\vec{r})$ at the Lagrange point between two atoms.
Our definition of bond order between atoms A and B is
\begin{eqnarray}
b_\varepsilon = \frac{\varepsilon^S_{\tau {\rm AB}}(\vec{r}_L)}{\varepsilon^S_{\tau {\rm HH}}(\vec{r}_L)}. \label{eq:be}
\end{eqnarray}
One should note we normalize by the value of a ${\rm H_2}$ molecule calculated at the same level of theory (including method and basis set).

\section{Results and discussion} \label{sec:results}

\subsection{Data set and computational details}
Our data set consists of 35 molecules (Table \ref{tab:GSTbond_1}, the leftmost column)
which may exist  in the gas phase of CVD process for GeSbTe alloy.
They could be formed by reactions among GST precursors and/or H$_2$ carrier gas,
and each of them has a bond between Ge, Sb or Te atoms.
Their geometries and coordinates are given in the supplementary material (Fig.~S1 and Table S1).
For the precursors, we assume $tert$-Butylgermanium (GeH$_3$(tBu), Fig.~\ref{fig:precursors}(a)) for the Ge precursor, 
triisopropylantimony (Sb(iPr)$_3$, Fig.~\ref{fig:precursors}(b)) for the Sb precursor, and 
diisopropyltellurium (Te(iPr)$_2$, Fig.~\ref{fig:precursors}(c)) for the Te precursor \cite{Machida2010}.
(Their coordinates are given in the supplementary material, Table S2.)
They are optimized by the DFT method based on 
the Lee-Yang-Parr gradient-corrected functional \cite{Lee1998,Miehlich1989}
with Becke's three hybrid parameters \cite{Becke1993} (B3LYP).
Threshold for maximum force is set to 0.000450 hartree/bohr.
The Dunning-Huzinaga double-zeta basis set \cite{Dunning1976}
with effective core potential by Hay and Wadt \cite{Hay1985a,Wadt1985,Hay1985b}
(LanL2DZ) are used for Ge, Sb and Te atoms.
6-31G(d) \cite{Hehre1972,Hariharan1974} basis set is used for C atoms.
D95(d,p) \cite{Dunning1976} basis set is employed for H atoms.
Total energies at 0\,K is obtained including zero-point energies with a scaling factor for B3LYP, 0.980 \cite{Bauschlicher1995}.

We also use Ge, Sb and Te crystal structures \cite{Singh1968,Schiferl1977,Cherin1967} as a reference data
(see Table S3 in the supplementary material for the details). 
We use the norm-conserving pseudopotentials of Troullier-Martins type \cite{Troullier1991} and 
the generalized-gradient approximation method by Perdew-Burke-Ernzerhof \cite{Perdew1996} for density
functional exchange-correlation interactions.
Kinetic energy cutoff of plane-wave expansion (k-point) is taken as 40.0 hartree ($2\times 2\times 2$ k-point set).

The electronic structures used in this paper are obtained by Gaussian 09 \cite{Gaussian09}
 for cluster models and by ABINIT \cite{ABINIT1,ABINIT2} for periodic models.
We use the QEDynamics package \cite{QEDynamics} developed in our group 
to compute the quantities described in the previous section
such as Eqs.~\eqref{eq:stress}, \eqref{eq:tension} and \eqref{eq:ked}.

\subsection{Electronic stress tensor and its eigenvalues}
We begin by briefly reviewing our past works on the electronic stress tensor analysis which are related to 
the present paper. 
First, it has been proposed that a covalent bond can be described 
by the eigenvalues and eigenvectors of the electronic stress tensor \cite{Tachibana2004}. 
In detail, the bonding region with covalency can be characterized and visualized by the ``spindle structure",
where the largest eigenvalue of the electronic stress tensor is positive and
 the corresponding eigenvectors form a bundle of flow lines that connects nuclei. 
As an example, we show a map of the largest eigenvalue of the electronic stress tensor including a region 
between C atoms of GeH$_3$(tBu) in Fig.~\ref{fig:GeH3tBu} (a), where we can find the spindle structure.
In passing, in Fig.~\ref{fig:GeH3tBu}  (b), we show the tension density and the Lagrange point for the same C-C bond.
Then, we have proposed that the negativity of the three eigenvalues of the stress tensor and their degeneracy,
which is the same pattern as liquid, 
can characterize some aspects of the metallic nature of chemical bonding \cite{Szarek2007,Ichikawa2012}. 
In Ref.~\cite{Ichikawa2012}, it has been shown that the three eigenvalues of the Li and Na clusters have almost same values while 
the hydrocarbon molecules have the largest eigenvalue much larger than the second largest eigenvalue, 
which has similar value to the smallest eigenvalue. 
In terms of the differential eigenvalues,
the Li and Na clusters have very small $\tau^{S 33}_{e}-\tau^{S 22}_{e}$ and $\tau^{S 22}_{e}-\tau^{S 11}_{e}$
which are much smaller than $\tau^{S 33}_{e}-\tau^{S 22}_{e}$ of hydrocarbons. 
The former degeneracy pattern indicates that the bonds are not directional while the latter indicates the clear directionality of the bonds,
reflecting the metallic nature of chemical bonding in the alkali metal clusters and the covalent nature of bonding in the hydrocarbon molecules. 

Now, let us move on to study the electronic stress tensor of the GST bonds.
We first show a map of $\tau^{S 33}_{e}$ and corresponding eigenvector on a plane which
contains a bond between GST atoms in Fig.\ref{fig:stressGST}. 
We find a Lagrange point between each bond and its position is marked in the figure. 
A common feature we see in all the six panels is that the eigenvectors form a pattern which connects GST nuclei.
However, not all of them are spindle structures. 
As for the Ge-Ge (panel (a)) and Ge-Sb (panel (d)) bonds, since they do not exhibit a positive $\tau^{S 33}_{e}$ region,
they are called to have pseudo-spindle structures \cite{Ichikawa2011c}.
As for the Te-Te (panel (c)) and Sb-Te (panel (f)) bonds, although we may say they have spindle structures,
the positive regions are not as conspicuous as the spindle structure of the C-C bond seen in Fig.~\ref{fig:GeH3tBu} (a).
As for the Sb-Sb (panel (b)) and Ge-Te (panel (e)) bonds, the positive $\tau^{S 33}_{e}$ regions are even smaller.
These results lead us to conclude that the GST atoms have the ability to form the spindle structure
in the order of Te $>$ Sb $>$ Ge. 
From our viewpoint that the spindle structure is the manifestation of the covalency of chemical bonding,
Te contributes to covalency more  than Sb or Ge, but less than C. 
The eigenvalue and eigenvector maps for the other GST molecules, which are found in the supplementary material
(Fig.~S2), also support this ordering. 

In Fig.~\ref{fig:dist_tau33}, we plot $\tau^{S 33}_{e}$ at the Lagrange point against the bond distance for
all the GST molecules. 
The original data are found in Table~\ref{tab:GSTbond_1}.
We see that $\tau^{S 33}_{e}>0$ for the Te-Te bonds and $\tau^{S 33}_{e}<0$ for the Ge-Ge bonds.
As for the Sb-Sb bonds, $\tau^{S 33}_{e}$ can be both positive and negative, and 
absolute values are smaller than those of the Ge-Ge and Te-Te bonds. 
$\tau^{S 33}_{e}$ of the Ge-Sb bonds exhibit intermediate values between the Ge-Ge and Sb-Sb bonds,
and similarly for the Ge-Te and Sb-Te bonds.
Therefore, Fig.~\ref{fig:dist_tau33} can be interpreted that the GST atoms contribute to the positivity of the 
$\tau^{S 33}_{e}$ at the Lagrange point in the order of Te $>$ Sb $>$ Ge,
which is consistent with the tendency to form the spindle structure as mentioned above. 

We next examine the differential eigenvalues of the electronic stress tensor at the Lagrange points.
In Fig.~\ref{fig:diffeig} (a), we show a scatter plot of $\tau^{S 33}_{e}-\tau^{S 22}_{e}$ and $\tau^{S 22}_{e}-\tau^{S 11}_{e}$
for GST bonds, and, in Fig.~\ref{fig:diffeig} (b),
we in addition plot points for the hydrocarbon molecules, Li clusters, and Na clusters,
which are taken from the data set studied in our previous paper \cite{Ichikawa2012}.
We see that both $\tau^{S 33}_{e}-\tau^{S 22}_{e}$ and $\tau^{S 22}_{e}-\tau^{S 11}_{e}$ 
of the Li and Na clusters are $O(10^{-4})$. 
As for the GST molecules, $\tau^{S 33}_{e}-\tau^{S 22}_{e}$ is $O(10^{-2})$, which is somewhat larger than
$\tau^{S 22}_{e}-\tau^{S 11}_{e}$, which is $O(10^{-4})$-$O(10^{-3})$.
As for the hydrocarbon molecules, $\tau^{S 33}_{e}-\tau^{S 22}_{e}$ is $O(10^{-1})$ and this is much larger than
$\tau^{S 22}_{e}-\tau^{S 11}_{e}$.
Thus, the degree of degeneracy can be summarized as ${\rm Li, Na} \ll {\rm GST} \ll {\rm h/c}$.
This is consistent with the usual classification of Ge, Sb and Te as metalloids,
which are placed in between metals and non-metals.
Further research may reveal that the electronic stress tensor density provides a new criterion to define metalloids based on the electronic structures. 

Incidentally, we compute the electronic stress tensor of Ge, Sb, and Te crystals using periodic models. 
We compute the electronic stress tensor at the midpoint of two nearest neighborhood atoms 
(it is the Lagrange point by symmetry).
The results are summarized in Table \ref{tab:GST_crystal} and plotted in Figs.~\ref{fig:dist_tau33} and \ref{fig:diffeig} (a).
We see in Fig.~\ref{fig:diffeig} (a) that all the crystal structures have similar values of $\tau^{S 33}_{e}-\tau^{S 22}_{e}$ and $\tau^{S 22}_{e}-\tau^{S 11}_{e}$ to those of the GST molecules.
This indicates that the degree of degeneracy does not differ much between the crystal structures and molecules. 
As for $\tau^{S 33}_{e}$, as shown in Fig.~\ref{fig:dist_tau33}, 
all the crystal structures exhibit negative values of about $-2 \times 10^{-3}$.
This is close to the $\tau^{S 33}_{e}$ of the Ge-Ge bonds in the molecules,
but not to those of the Sb-Sb or Te-Te bonds, which, respectively, are negative with absolute values of $O(10^{-5})$ or 
positive with values of $O(10^{-3})$.
Thus, from the viewpoint of the electronic stress tensor, 
the covalency of the chemical bonding involving Sb and Te atoms in the GST molecules does not
appear in their crystal structures and some metallicity is manifested. 
In other words, while Ge, Sb and Te do not exhibit covalency in their crystal structure,
Sb and Te show some covalency in their molecular structure.

\subsection{Bond order and force constant}

We briefly review our past works \cite{Szarek2007,Szarek2008,Szarek2009,Ichikawa2011c} on our bond order $b_\varepsilon$ (Eq.~\eqref{eq:be}), which
is defined using concepts based on the Rigged QED.
First, it has been pointed out \cite{Szarek2007} that $b_\varepsilon$ of a single, double, and triple bond 
between carbon atoms in hydrocarbons is close to 1, 2 and 3 respectively, consistent with a conventional bond order. 
Also, $b_\varepsilon$ of the C-C bond in a benzene molecule is close to 1.5. 
However, it has been also reported that $b_\varepsilon$ of some diatomic molecules overestimate or underestimate the conventional bond order, e.g., $b_\varepsilon$ of N$_2$ is 7.462.
Then, in Ref.~\cite{Szarek2008}, the correlation between $b_\varepsilon$ and bond distance $r_e$ has been investigated using various simple organic compounds. 
By comparing with other bond orders proposed in the literature, $b_\varepsilon$ is found to have a comparable or better correlation. 
This has been also shown using hydrogenated Pt \cite{Szarek2009} and Pd \cite{Ichikawa2011c} clusters. 
We, however, have not investigated a correlation between $b_\varepsilon$ and a quantity related to energy. 
Therefore, we here compute the bond dissociation energy $B$ and force constant $k$ of the GST bonds, and investigate the correlations with $b_\varepsilon$.

The results are summarized in Table \ref{tab:GSTbond_2}. 
$k$ is calculated based on the directional derivative of a total energy $E$ of a molecule with respect to the bond direction $\vec{s}$. Namely, it is computed using force constant matrix $\frac{\partial^2 E}{\partial x^i \partial x^j}$ as
\begin{eqnarray}
k = - \frac{\partial^2 E}{\partial \vec{s} \partial \vec{s}} = - \sum_{i,j=1}^3 s^i  \frac{\partial^2 E}{\partial x^i \partial x^j} s^j. \label{eq:k}
\end{eqnarray}
The scatter plots of $b_\varepsilon$ versus $r_e$, $B$, and $k$ are shown in Fig.~\ref{fig:be}.
We can see that $b_\varepsilon$ is negatively correlated with $r_e$ and positively correlated with $k$ and $B$,
consistently with the usual notion of bond order.
In fact, their correlation coefficients are $-0.968$, 0.570, and 0.910, respectively for $r_e$, $B$, and $k$.
Since we need to know the energy when two fragments of a molecule are infinitely apart to compute $B$,
it is reasonable not to find a strong correlation between $b_\varepsilon$ and $B$. 
Note that $b_\varepsilon$ is defined only using the quantities at the equilibrium structure. 
On the other hand, the reason why the correlation with $k$ is very strong may be understood as follows. 

The meaning of $k$, often called a spring constant, is how much energy we need to 
stretch the bond by an infinitesimal distance.
It can be expressed as $k \delta^2/2$, where $\delta$ denotes the infinitesimal displacement.
We may interpret this energy using our energy density $\varepsilon_\tau^S(\vec{r})$ (Eq.~\eqref{eq:ed}) 
and the ``Lagrange surface" \cite{Tachibana2010,Tachibana2013} which is defined as a separatrix in the tension density $\vec{\tau}^S(\vec{r})$ (Eq.~\eqref{eq:tension}). 
The vector field of $\vec{\tau}^S(\vec{r})$ generally has a pattern in which vectors originate from atomic nuclei,
and they collide to form separatrix. See Figs.~\ref{fig:GeH3tBu}~(b) or \ref{fig:GeH3SbH2}~(a).
We call this separatrix the Lagrange surface and regard it as the boundary of atoms in a molecule.
In Fig.~\ref{fig:GeH3SbH2} (a), we show some examples of the Lagrange surface in a GeH$_3$-SbH$_2$ molecule. 
When we move apart two atoms bounded by the Lagrange surface for a small distance, 
it may be reasonable to suppose the required energy to be proportional to the energy stored in the Lagrange surface,
that is, $\int_{\cal S} \varepsilon_\tau^S(\vec{\sigma}) d^2\sigma$, where the integration is taken over the Lagrange surface ${\cal S}$. 
To support this idea, we compute an alternative bond order \cite{Ichikawa2011b,Ichikawa2011c} defined as
\begin{eqnarray}
b_{\varepsilon(S)} =
\frac{ \int_{{\cal S}_{\rm AB}}  \varepsilon^S_\tau(\vec{\sigma})d^2\sigma }{ \int_{{\cal S}_{\rm HH}} \varepsilon^S_\tau(\vec{\sigma}) d^2\sigma},
\label{eq:beS}
\end{eqnarray}
where ${\cal S}_{\rm AB}$ denotes the Lagrange surface between atoms A and B,
and investigate the correlation with $k$ (Fig.~S3 (c)).
The correlation coefficient is found to be 0.887, 
and the linear fit $y = a x + b$, where $x$ and $y$, respectively, represent $k$ and $b_{\varepsilon(S)}$, gives $a = 5.54$ and $b = 5.62 \times 10^{-3}$. (The value of $b_{\varepsilon(S)}$ for each bond is summarized in Table \ref{tab:GSTbond_2}.)
This suggests that $k \propto \int_{\cal S} \varepsilon_\tau^S(\vec{\sigma}) d^2\sigma$ holds to a large extent.
As for the linear fit for $b_\varepsilon$ against $k$,  we obtain $a=1.63$ and $b=-7.54 \times 10^{-3}$.
Since $b$ is close to zero for both cases, we can approximately consider $b_{\varepsilon(S)} \propto b_\varepsilon$ with a proportional constant of $3.4$ (the ratio of $a$'s) for these GST bonds. 
This relation between $b_{\varepsilon(S)}$ and $b_\varepsilon$ can be confirmed directly by computing
$b_{\varepsilon(S)}/b_\varepsilon$ for each GST bond which is shown in Fig.~\ref{fig:beSoverbe}. 
The average of $b_{\varepsilon(S)}/b_\varepsilon$ over 35 bonds is $3.74$ with the standard deviation of $0.21$.
Since the standard deviation is relatively small (5.6\% of the average), we may regard $b_{\varepsilon(S)} \propto b_\varepsilon$ for these bonds. 
The values of the proportional constant derived in two ways are consistent. 

This relation, that $b_{\varepsilon(S)}$ of the GST bonds is approximately obtained by multiplying $b_\varepsilon$ by a common factor, holds if 
the integration of $\varepsilon^S_\tau(\vec{r})$ over their Lagrange surface is well approximated by $\varepsilon^S_\tau(\vec{r}_L)$ multiplied by a common factor. 
Such an assumption can be valid, if the Lagrange surfaces of the GST bonds are flat
and the energy density distributions on them are expressed by Gaussian functions with a common value of the exponent. 
We can see this is roughly true as follows.
First, the flatness can be checked by the visual inspection of the Lagrange surface (Fig.~\ref{fig:GeH3SbH2} (a) and Fig.~S2 in the supplementary material). 
Then, we plot $\varepsilon^S_\tau(\vec{r})$ against the distance from $\vec{r}_L$ for the points which constitute the Lagrange surface. 
The example of this plot for the GeH$_3$-SbH$_2$ molecule is found in Fig.~\ref{fig:GeH3SbH2} (b).
 Actually, we plot $\varepsilon^S_\tau(\vec{r})$ divided by the energy density of a hydrogen molecule at its Lagrange point 
so that the value at $\vec{r}_L$ ({\it i.e.} $|\vec{r}-\vec{r}_L|=0$) becomes $b_\varepsilon$.
In the figure, we also plot the result of the fit to the Gaussian function $b_\varepsilon \exp \left\{ -\alpha |\vec{r}-\vec{r}_L|^2 \right\}$ where the exponent $\alpha$ is the fitting parameter. 
The figure shows that $\varepsilon^S_\tau(\vec{r})/\varepsilon^S_{\tau {\rm HH}}(\vec{r}_L)$ is well fitted by the Gaussian function with $\alpha = 1.32$.
We perform such a fit to all the GST bonds (Figures similar to Fig.~\ref{fig:GeH3SbH2} (b) are plotted in Fig.~S4. The values of $\alpha$ are summarized in Table \ref{tab:GSTbond_2} and Fig.~S5) and the average and standard deviation of the exponent are computed to be 1.31 and  $9.40 \times 10^{-2}$, respectively. 
Therefore, as for the GST bonds, we can say that the energy density distribution over the Lagrange surface is expressed by the Gaussian functions with a common exponent, 
which leads to $\int_{\cal S} \varepsilon_\tau^S(\vec{\sigma}) d^2\sigma \propto \varepsilon^S_\tau(\vec{r}_L)$.

As a further test for the idea that  $b_{\varepsilon(S)}$ is proportional to the force constant, we examine homonuclear diatomic molecules,
from H$_2$ to I$_2$ excluding the group 18 elements. 
We use experimental values for $r_e$, $B$, and $k$ as is summarized in Table \ref{tab:dm}.
$B$ is the energy of the ground state atomic products relative to the lowest existing level of the molecule, often denoted by $D_0$.
$k$ is converted from the frequency $\omega$ (which is also shown in the table) via $k = M \omega^2/2$, where $M$ is the atomic mass.
Note that some of the experimental values are not available or controversial and we use computational values in such cases. 
We adopt computational values for $r_e$ and $\omega$ of Ga$_2$ from Ref.~\cite{Das1997}, and $r_e$ and $\omega$ of In$_2$ from Ref.~\cite{Balasubramanian1988}.
$b_{\varepsilon}$ and $b_{\varepsilon(S)}$ are computed with the same setups as the GST bonds, and shown in Table \ref{tab:dm}.
We first study how much $b_{\varepsilon}$ is correlated with $r_e$, $B$, and $k$.
The correlation coefficients are found to be $-0.583$, 0.834, and 0.978, respectively for $r_e$, $B$, and $k$,
showing very strong correlation between $b_{\varepsilon}$ and $k$ as is the case with the GST bonds. 
As for $b_{\varepsilon(S)}$, the correlation with $k$ is also very strong, with the correlation coefficient 0.990.
The scatter plots of $k$ versus $b_{\varepsilon}$ and $b_{\varepsilon(S)}$ are shown in Fig~\ref{fig:be_DM} (a) and (b) respectively, and we can see these strong correlations (similar scatter plots for $r_e$ and $B$ are shown in Figs.~S6 and S7). 
However, one may notice a difference between Fig~\ref{fig:be_DM} (a) and (b) that the panel (b) shows more proportionality than the panel (a).
In fact, while the linear fit $b_{\varepsilon} = a_1 k + b_1$ gives $a_1 = 3.12$ and $b_1 = -0.23$,
$b_{\varepsilon(S)} = a_2 k + b_2$ gives $a_2 = 3.64$ and $b_2 = 5.87 \times 10^{-3}$. 
Since $b_1$ is not negligible, $b_{\varepsilon}$ is hardly considered to be proportional to $k$, whereas very small value of $b_2$ implies that 
$b_{\varepsilon(S)} \propto k$ holds to a large extent.
Therefore, in general, $b_{\varepsilon(S)}$ is a better descriptor of a chemical bond than $b_{\varepsilon}$. 
The drawback is that, since the computation of $b_{\varepsilon(S)}$ involves the integration over the Lagrange surface,
it costs much greater than that of $b_{\varepsilon}$. 

Finally, some comments on the correlations of $r_e$ and $B$ with our bond orders are in order. 
First, the correlation coefficient between $r_e$ and $b_\varepsilon$ is $-0.968$ for the GST molecules and 
$-0.583$ for the homonuclear diatomic molecules. (As for $b_{\varepsilon(S)}$, they are $-0.970$ and $-0.651$, respectively.)
The difference can be attributed to the fact that the GST data consists of similar chemical bonds formed by Ge, Sb, and Te,
whereas the homonuclear diatomic molecule data contains various types of elements. 
It has been found that $b_\varepsilon$ correlates well with $r_e$, as mentioned in the beginning of this subsection, 
but it has been also shown that the slope on the $r_e$-$b_\varepsilon$ plane depends on the type of chemical bonding.
For example, in Ref.~\cite{Ichikawa2011c}, the Pd-Pd bonds and Pd-H bonds are found to have different slopes. 
Also, among the Pd-H bonds, the bonds with shorter bond length ($\lesssim 1.9$\,{\AA})
have steeper slope than those with longer bond length. 
Therefore, when a data contains different types of chemical bonding, even though $r_e$ and $b_\varepsilon$ exhibit negative correlation,
the correlation coefficient tends to be somewhat away from $-1$.
Second, the correlation coefficient between $B$ and $b_\varepsilon$ is $0.570$ for the GST data and 
$0.834$ for the homonuclear diatomic molecules data. (As for $b_{\varepsilon(S)}$, they are $0.569$ and $0.906$ respectively.)
There is again some notable difference between the two data sets. 
Since the correlation coefficients between $b_\varepsilon$ and $k$ are close to 1 for both data sets, 
the degrees of correlation between $B$ and $b_\varepsilon$ are roughly equal to those between $B$ and $k$.
Although $B$ and $k$ are likely to be positively correlated, since they are not linear to each other in general,
we do not expect a very strong correlation between $B$ and $k$. 
Then, the relatively low value of correlation coefficient of the GST data is reasonable while that of the homonuclear diatomic molecules data is unexpectedly high. The latter may be understood by a universality of potential energy curve for diatomic molecules (e.g. \cite{Sutherland1940,Graves1985,Xie2006}), which has been studied for a long time.
It is, however, an empirical relation at this stage and the discussion of its relevance here is beyond the scope of this paper.

\section{Conclusions} \label{sec:conclusion}

We have analyzed the electronic structure of 35 molecules which may exist in gas phase of CVD 
process for GeSbTe alloy using the electronic stress tensor,
with special focus on the chemical bonds between Ge, Sb and Te atoms.
Our study consists of two parts. 
First, we have studied the pattern of the eigenvalues and eigenvectors of the
electronic stress tensor density of the GST bonds. 
Next, we have computed the bond order which is defined by the stress-tensor-based energy density
for the GST bonds, and have investigated the correlation with the energy-related quantities
such as the bond dissociation energy and force constant. 

In the first part,
we have found that, from the viewpoint of the electronic stress tensor density, 
GST bonds exhibit intermediate properties between alkali metals and hydrocarbon molecules.
This is illustrated by the sign and degeneracy pattern of the three eigenvalues of the electronic stress tensor density
at the Lagrange point between two atoms.  
In our previous studies, we have pointed out that the negative and degenerate eigenvalues, which indicate
a lack of directionality with the compressive stress, characterize some aspects of metallicity of chemical bonding. 
By contrast, the positive largest eigenvalue and much smaller negative other two eigenvalues,
which indicates a solid directionality with the tensile stress in that direction, characterize covalency.
The former pattern has been typically found in the alkali metals and the latter in the hydrocarbon molecules.
Our results in the present paper suggest that the GST bonds can be located in between metallic bonding and covalent bonding 
in terms of their electronic stress tensor. 
This is consistent with the usual classification of Ge, Sb and Te as metalloids, which have
intermediate properties between metals and nonmetals.

In the second part,
we have found that the correlation of our bond order with the bond dissociation energy is not so strong, while one 
with the force constant is very strong. 
We have interpreted this results in terms of the energy density
on the ``Lagrange surface", which is considered to define the boundary surface of atoms in a molecule
in the framework of the electronic stress tensor analysis. 
In this study, we have found that both of our definitions of bond order $b_\varepsilon$ and $b_{\varepsilon (S)}$,
where the former uses the energy density at the Lagrange point while the latter involves the integration over
the Lagrange surface, have strong correlation with the force constant. 
We have argued that, if the interpretation above is correct, $b_{\varepsilon (S)}$ is the one which is more directly
 connected with the force constant, and the strong correlation of the force constant with $b_\varepsilon$ follows from that with $b_{\varepsilon (S)}$.
In fact, we have shown that $b_{\varepsilon (S)}/b_\varepsilon$ does not vary much among the GST bonds,
which originate from the fact that 
the energy distributions on the Lagrange surface of the GST bonds can be well expressed 
by Gaussian functions centered at the Lagrange point and with a common value of the exponent.
As the results of this study, it is hinted that 
 the stress-tensor-based energy density can be related not only to the total energy but also
 to the force constant by combining with another Rigged QED concept, the Lagrange surface.

In our future work, regarding the first part, 
we wish to apply the electronic stress tensor analysis
to the other elements which are conventionally classified as metalloids, B, Si and As, 
to see whether it can provide a criterion to define metalloids. 
For that purpose, it is also necessary to extend the analysis to the elements nearby metalloids in the periodic table.
We also wish to apply the electronic stress tensor analysis to 
transition metals, ionic bonds, hypervalency, and so on, to strengthen its usefulness. 
It would enable us to deepen our understanding of the nature of chemical bonding. 
As for the second part, 
a further direction of the study will be to provide more evidence for our results by using other types of molecules. 
If the relation concerning the force constant, which is an experimental observable derived from a vibrational spectrum,
 is established in general, it would help to solidify such ideas as the stress-tensor-based energy density and Lagrange surface.
In the end, we would like to emphasize that our studies are based on the quantities defined at each point in space, 
which originate from the quantum field theoretic consideration and not from the electron density.
We believe that our method will lead us to new and beneficial views on chemical systems. 

\section*{Acknowledgments}  

Theoretical calculations were partly performed using Research Center for
 Computational Science, Okazaki, Japan.
This work is supported by Grant-in-Aid for Scientific research 
(No.~25410012) from the Ministry of Education, Culture, Sports, Science and Technology, Japan.
This research work is supported by the national project ``Green Nanoelectronics".
Y. I. is supported by the Sasakawa Scientific Research Grant from the
Japan Science Society.


\renewcommand{\baselinestretch}{1}

\newpage

\begin{table}[htbp]
\begin{center}
\caption{Data for the bonds between Ge, Sb, and Te atoms in the GST molecules.
$r_e$ is the bond distance. $\tau_e^{S33}$, $\tau_e^{S22}$, and $\tau_e^{S11}$ are three eigenvalues of 
the electronic stress tensor density at the Lagrange point.}
\vspace{5mm}
\begin{tabular}{l r r r r}
\hline
\hline
Molecule&
$r_e$\,[{\AA}] &
$\tau_e^{S33} (\times 10^3) $&
$\tau_e^{S22}(\times 10^3) $&
$\tau_e^{S11}(\times 10^3) $\\

\hline
${\rm GeH_2}$-${\rm GeH_2}$&
2.343&$-7.533$&$-18.29$&$-20.08$\\
${\rm GeH_2(tBu)}$-${\rm GeH_2(tBu)}$&
2.489&$-1.811$&$-16.21$&$-16.31$\\
${\rm GeH_3}$-${\rm GeH_2}$&
2.489&$-1.768$&$-15.71$&$-15.95$\\
${\rm GeH_3}$-${\rm GeH_2(tBu)}$&
2.486&$-1.841$&$-16.23$&$-16.28$\\
${\rm GeH_3}$-${\rm GeH_3}$&
2.483&$-1.861$&$-16.25$&$-16.25$\\

\hline
${\rm SbH_2}$-${\rm SbH_2}$&
2.948&$\hspace{2.8mm}0.136$&$-9.121$&$-9.562$\\
${\rm SbH(iPr)}$-${\rm SbH_2}$&
2.946&$\hspace{2.8mm}0.053$&$-9.204$&$-9.654$\\
${\rm SbH(iPr)}$-${\rm SbH(iPr)}$&
2.937&$-$$0.061$&$-9.392$&$-9.891$\\
${\rm Sb(iPr)_2}$-${\rm SbH_2}$&
2.944&$\hspace{2.8mm}0.022$&$-9.335$&$-9.724$\\
${\rm Sb(iPr)_2}$-${\rm SbH(iPr)}$&
2.939&$-0.055$&$-9.451$&$-9.897$\\
${\rm Sb(iPr)_2}$-${\rm Sb(iPr)_2}$&
2.954&$\hspace{2.8mm}0.213$&$-9.226$&$-9.752$\\

\hline
${\rm TeH}$-${\rm TeH}$&
2.854&$\hspace{2.8mm}1.897$&$-12.43$&$-12.45$\\
${\rm Te(iPr)}$-${\rm TeH}$&
2.844&$\hspace{2.8mm}1.514$&$-12.69$&$-12.82$\\
${\rm Te(iPr)}$-${\rm Te(iPr)}$&
2.831&$\hspace{2.8mm}1.139$&$-13.03$&$-13.24$\\

\hline
${\rm GeH_2}$-${\rm SbH_2}$&
2.703&$-0.940$&$-12.04$&$-12.61$\\
${\rm GeH_2}$-${\rm SbH(iPr)}$&
2.696&$-1.102$&$-12.15$&$-12.83$\\
${\rm GeH_2}$-${\rm Sb(iPr)_2}$&
2.692&$-1.218$&$-12.25$&$-12.99$\\
${\rm GeH_2(tBu)}$-${\rm SbH_2}$&
2.703&$-0.860$&$-12.49$&$-12.86$\\
${\rm GeH_2(tBu)}$-${\rm SbH(iPr)}$&
2.702&$-0.894$&$-12.59$&$-12.97$\\
${\rm GeH_2(tBu)}$-${\rm Sb(iPr)_2}$&
2.704&$-0.869$&$-12.58$&$-13.01$\\
${\rm GeH_3}$-${\rm SbH_2}$&
2.700&$-0.811$&$-12.52$&$-12.82$\\
${\rm GeH_3}$-${\rm SbH(iPr)}$&
2.694&$-0.995$&$-12.70$&$-13.04$\\
${\rm GeH_3}$-${\rm Sb(iPr)_2}$&
2.696&$-0.987$&$-12.69$&$-13.07$\\

\hline
${\rm GeH_2(tBu)}$-${\rm Te(iPr)}$&
2.654&$\hspace{2.8mm}0.059$&$-14.36$&$-15.28$\\
${\rm GeH_2(tBu)}$-${\rm TeH}$&
2.658&$\hspace{2.8mm}0.134$&$-14.05$&$-15.15$\\
${\rm GeH_2}$-${\rm TeH}$&
2.653&$-0.649$&$-13.82$&$-15.08$\\
${\rm GeH_2}$-${\rm Te(iPr)}$&
2.637&$-1.078$&$-14.32$&$-15.57$\\
${\rm GeH_3}$-${\rm TeH}$&
2.649&$\hspace{2.8mm}0.209$&$-14.28$&$-15.30$\\
${\rm GeH_3}$-${\rm Te(iPr)}$&
2.636&$-0.183$&$-14.76$&$-15.75$\\

\hline
${\rm SbH_2}$-${\rm TeH}$&
2.892&$\hspace{2.8mm}0.838$&$-10.60$&$-11.12$\\
${\rm SbH_2}$-${\rm Te(iPr)}$&
2.874&$\hspace{2.8mm}0.343$&$-11.05$&$-11.52$\\
${\rm SbH(iPr)}$-${\rm TeH}$&
2.890&$\hspace{2.8mm}0.627$&$-10.59$&$-11.28$\\
${\rm SbH(iPr)}$-${\rm Te(iPr)}$&
2.878&$\hspace{2.8mm}0.441$&$-10.97$&$-11.58$\\
${\rm Sb(iPr)_2}$-${\rm TeH}$&
2.892&$\hspace{2.8mm}0.410$&$-10.66$&$-11.20$\\
${\rm Sb(iPr)_2}$-${\rm Te(iPr)}$&
2.879&$\hspace{2.8mm}0.338$&$-11.05$&$-11.53$\\

\hline
\end{tabular}
\label{tab:GSTbond_1}
\end{center}
\end{table}

\begin{table}[htbp]
\begin{center}
\caption{Data for the bonds in the crystal structure of Ge, Sb, and Te.
$r_e$ is the distance between the nearest neighborhood atoms.
$\tau_e^{S33}$, $\tau_e^{S22}$, and $\tau_e^{S11}$ are three eigenvalues of 
the electronic stress tensor density at the Lagrange point.}
\vspace{5mm}
\begin{tabular}{c r r r r}
\hline
\hline
Crystal &
$r_e$\,[{\AA}] &
$\tau_e^{S33}$ ($\times 10^3$) &
$\tau_e^{S22}$ ($\times 10^3$) &
$\tau_e^{S11}$ ($\times 10^3$) \\

\hline
Ge 
&2.450& $-2.612$ &$-19.52$ &$-19.52$ \\
Sb
&2.907& $-1.981$ &$-12.12$&$-12.23$\\
Te 
&2.835&$-1.712$&$-17.46$&$-17.81$\\
\hline
\end{tabular}
\label{tab:GST_crystal}
\end{center}
\end{table}

\begin{table}[htbp]
\begin{center}
\caption{Data for the bonds between Ge, Sb, and Te atoms in the GST molecules.
$B$ and $k$ are respectively the bond dissociation energy and force constant.
$b_\varepsilon$ and $b_{\varepsilon (S)}$ are our bond orders (Eqs.~\eqref{eq:be} and \eqref{eq:beS}).
$\alpha$ is a fitting parameter for the energy density distribution on the Lagrange surface (see the text for the details).}
\vspace{5mm}
\begin{tabular}{l r r r r r}
\hline
\hline
\multicolumn{1}{l}{Molecule}&
\multicolumn{1}{c}{$B$ ${\rm [kcal/mol]}$}&
\multicolumn{1}{c}{$k$ ${\rm [a.u.]}$}&
\multicolumn{1}{c}{$b_\varepsilon$}&
\multicolumn{1}{c}{$b_{\varepsilon (S)}$}&
\multicolumn{1}{c}{$\alpha$}\\

\hline
${\rm GeH_2}$-${\rm GeH_2}$&
40.40&0.122&0.179&0.648&1.561\\
${\rm GeH_2(tBu)}$-${\rm GeH_2(tBu)}$&
59.79&0.080&0.132&0.485&1.443\\
${\rm GeH_3}$-${\rm GeH_2}$&
40.30&0.079&0.128&0.473&1.444\\
${\rm GeH_3}$-${\rm GeH_2(tBu)}$&
59.86&0.082&0.132&0.484&1.448\\
${\rm GeH_3}$-${\rm GeH_3}$&
59.58&0.083&0.132&0.482&1.455\\

\hline
${\rm SbH_2}$-${\rm SbH_2}$&
32.21&  0.058& 	0.071&0.284&1.186\\
${\rm SbH(iPr)}$-${\rm SbH_2}$&
32.62& 	0.056& 	0.072&0.291&1.175\\
${\rm SbH(iPr)}$-${\rm SbH(iPr)}$&
33.09& 	0.060& 	0.074&0.301&1.177\\
${\rm Sb(iPr)_2}$-${\rm SbH_2}$&
32.72& 	0.056& 	0.073&0.297&1.166\\
${\rm Sb(iPr)_2}$-${\rm SbH(iPr)}$&
33.19& 	0.056& 	0.075&0.303&1.174\\
${\rm Sb(iPr)_2}$-${\rm Sb(iPr)_2}$&
31.05& 	0.056& 	0.072&0.294&1.164\\

\hline
${\rm TeH}$-${\rm TeH}$&
33.24& 	0.064& 	0.088&0.288&1.279\\
${\rm Te(iPr)}$-${\rm TeH}$&
34.86& 	0.063& 	0.092&0.307&1.278\\
${\rm Te(iPr)}$-${\rm Te(iPr)}$&
35.32& 	0.066& 	0.097&0.327&1.285\\

\hline
${\rm GeH_2}$-${\rm SbH_2}$&
28.06& 	0.069& 	0.098&0.378&1.323\\
${\rm GeH_2}$-${\rm SbH(iPr)}$&
28.94& 	0.068& 	0.100&0.390&1.316\\
${\rm GeH_2}$-${\rm Sb(iPr)_2}$&
30.10& 	0.069& 	0.102&0.396&1.317\\
${\rm GeH_2(tBu)}$-${\rm SbH_2}$&
45.95& 	0.060& 	0.101&0.389&1.314\\
${\rm GeH_2(tBu)}$-${\rm SbH(iPr)}$&
45.75& 	0.060& 	0.102&0.395&1.309\\
${\rm GeH_2(tBu)}$-${\rm Sb(iPr)_2}$&
46.22& 	0.062& 	0.102&0.398&1.298\\
${\rm GeH_3}$-${\rm SbH_2}$&
45.13& 	0.063& 	0.100&0.385&1.323\\
${\rm GeH_3}$-${\rm SbH(iPr)}$&
45.72& 	0.064& 	0.103&0.399&1.318\\
${\rm GeH_3}$-${\rm Sb(iPr)_2}$&
45.95& 	0.065& 	0.103&0.401&1.309\\

\hline
${\rm GeH_2(tBu)}$-${\rm Te(iPr)}$&
49.93& 	0.064&	0.114&0.404&1.357\\
${\rm GeH_2(tBu)}$-${\rm TeH}$&
52.67&	0.064&	0.112&0.415&1.355\\
${\rm GeH_2}$-${\rm TeH}$&
33.96& 	0.075&	0.114&0.403&1.393\\
${\rm GeH_2}$-${\rm Te(iPr)}$&
33.70& 	0.077&	0.119&0.428&1.391\\
${\rm GeH_3}$-${\rm TeH}$&
50.13& 	0.069&	0.113&0.404&1.375\\
${\rm GeH_3}$-${\rm Te(iPr)}$&
49.04& 	0.071&	0.118&0.427&1.375\\

\hline
${\rm SbH_2}$-${\rm TeH}$&
35.71&	0.061&	0.080&0.289&1.246\\
${\rm SbH_2}$-${\rm Te(iPr)}$&
34.54&	0.052&	0.085&0.310&1.260\\
${\rm SbH(iPr)}$-${\rm TeH}$&
38.52&	0.061&	0.082&0.297&1.236\\
${\rm SbH(iPr)}$-${\rm Te(iPr)}$&
37.04&	0.061&	0.085&0.297&1.233\\
${\rm Sb(iPr)_2}$-${\rm TeH}$&
39.79&	0.048&	0.082&0.301&1.239\\
${\rm Sb(iPr)_2}$-${\rm Te(iPr)}$&
38.36&	0.060&	0.085&0.319&1.233\\

\hline
\end{tabular}
\label{tab:GSTbond_2}
\end{center}
\end{table}

\begin{table}[htbp]
\begin{center}
\caption{Data for the homonuclear diatomic molecules. $r_e$, $\omega$, and $B$ are respectively the bond distance, frequency, and dissociation energy, whose values are taken from the references. The force constant $k$ is converted from $\omega$. Details are found in the text. $b_\varepsilon$ and $b_{\varepsilon (S)}$ are our bond orders (Eqs.~\eqref{eq:be} and \eqref{eq:beS}).}
\vspace{5mm}
\begin{tabular}{c | r r r r r | c c }
\hline
\hline
Molecule&
$r_e$\,[{\AA}] &
$\omega$\,[cm$^{-1}$]& $k$\,[a.u.] &
$B$\,[kcal/mol]&  
References & ~~~$b_\varepsilon$~~~  & ~~$b_{\varepsilon(S)}$~~ \\
\hline
H$_2$ &  0.74144 & 4401.21  & 0.367   & 103.26 &  \cite{HH1979} & 1.000 & 1.000 \\
\hline
Li$_2$ &  2.6729  & 351.43   &   0.016          & 24.2   & \cite{HH1979} &  0.013& 0.112\\
Be$_2$ & 2.453  & 270.7  & 0.012    & 2.308   & \cite{Merritt2009,Heaven2011} & 0.035 &0.165\\
B$_2$ & 1.590 & 1051.3 & 0.230   & 69 &  \cite{HH1979} & 0.452  &1.205\\
C$_2$ & 1.2425 & 1854.71 & 0.781   & 143  &  \cite{HH1979} & 2.349  &3.542\\
N$_2$ &  1.09769 & 2358.57 & 1.474   & 225.0 & \cite{HH1979} & 5.849  &6.437\\
O$_2$ &  1.20752 & 1580.19 & 0.756   & 118.0 &  \cite{HH1979} & 3.229 &4.079 \\
F$_2$ & 1.41193 & 916.64 & 0.302   & 36.94 &  \cite{HH1979} & 1.303 & 1.425\\
\hline
Na$_2$ & 3.0789 & 159.125 & 0.011   & 16.6 &  \cite{HH1979} & 0.006 &0.065\\
Mg$_2$ & 3.891 & 51.12 & 0.001   & 1.16 &  \cite{HH1979} & 0.004 & 0.022\\
Al$_2$  &  2.466 & 350.01 &0.063    & 37 &  \cite{HH1979} & 0.074 & 0.374\\
Si$_2$ &  2.246 & 510.98& 0.138    & 74.0 &  \cite{HH1979} & 0.245 & 0.802\\
P$_2$ & 1.8934 & 780.77 & 0.358   & 116.1 & \cite{HH1979} & 0.650 & 1.701\\
S$_2$ & 1.8892 & 725.65&  0.319   & 100.76 & \cite{HH1979} & 0.701 & 1.513\\
Cl$_2$ & 1.988 & 559.78 & 0.208    & 57.1742 & \cite{HH1979}& 0.421  &0.645 \\
\hline
K$_2$ & 3.9051 & 92.021 & 0.006   & 11.9 & \cite{HH1979} & 0.004 & 0.044\\
Ca$_2$ & 4.2773 & 64.93 &  0.003  & 3.0 & \cite{HH1979} & 0.004 & 0.038\\
Ga$_2$ & 2.746  &  162 & 0.034   &  26.36 & \cite{Das1997,Shim1991} & 0.042 & 0.133\\
Ge$_2$ & 2.3680 & 287.9 &  0.116  & 65 & \cite{Hostutler2002,HH1979}& 0.122 & 0.478\\ 
As$_2$ & 2.1026 & 429.55 &  0.262  & 91.3 & \cite{HH1979} & 0.423 & 1.286\\
Se$_2$ & 2.166 & 385.303 &  0.225  & 78.66 & \cite{HH1979} & 0.379 & 1.015\\
Br$_2$ & 2.2811 & 325.321 &  0.158  & 45.444 & \cite{HH1979} & 0.246 & 0.540\\
\hline
Rb$_2$ & 4.2099  & 57.781 & 0.005   & 11 & \cite{Amiot1985,HH1979} & 0.004 & 0.042\\
Sr$_2$ & 4.67174 & 40.32831 & 0.003   & 3.036 & \cite{Stein2008} & 0.003  & 0.031\\
In$_2$ & 3.14 & 111 & 0.027   & 17.8 & \cite{Balasubramanian1988,Balducci1998} & 0.034 &0.170\\
Sn$_2$ & 2.746 & 189.74 & 0.082   & 43.8300 & \cite{Bondybey1983,Pak1988} & 0.108 & 0.499\\
Sb$_2$ & 2.476 & 269.623 & 0.167   & 69.0672 & \cite{Sontag1982} & 0.220 & 0.869\\
Te$_2$ & 2.5574 & 247.07 &  0.150  & 61.73 & \cite{HH1979} & 0.195 & 0.678\\
I$_2$ & 2.666 & 214.50 &  0.111  & 35.5672 & \cite{HH1979} & 0.137 & 0.385\\
\hline
\end{tabular}
\label{tab:dm}
\end{center}
\end{table}

\ifFIG

\clearpage

\begin{figure}
\begin{center}
\includegraphics[width=15cm]{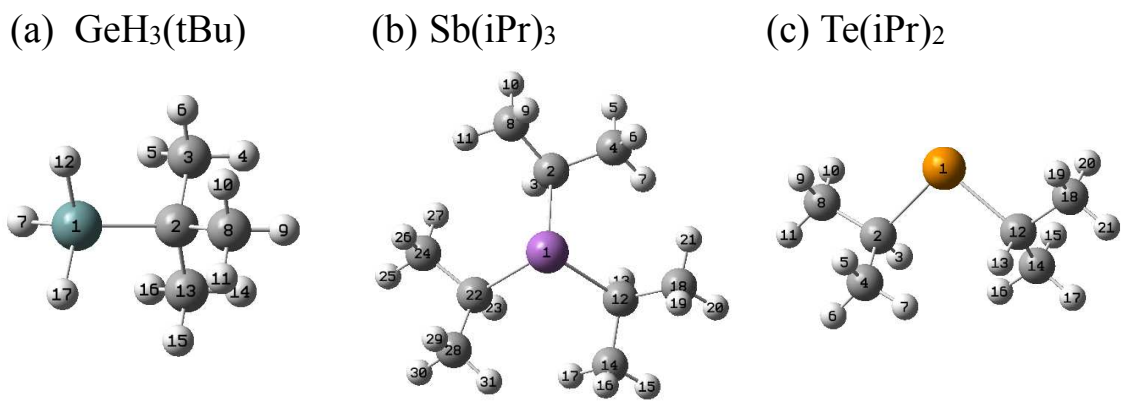}
\end{center}
\caption{The optimized structures of the precursors: (a) GeH$_3$(tBu) ($C_{3v}$), (b) Sb(iPr)$_3$ ($C_3$), and (c) Te(iPr)$_2$ ($C_2$). }
\label{fig:precursors}
\end{figure}

\begin{figure}
\begin{center}
\includegraphics[width=15cm]{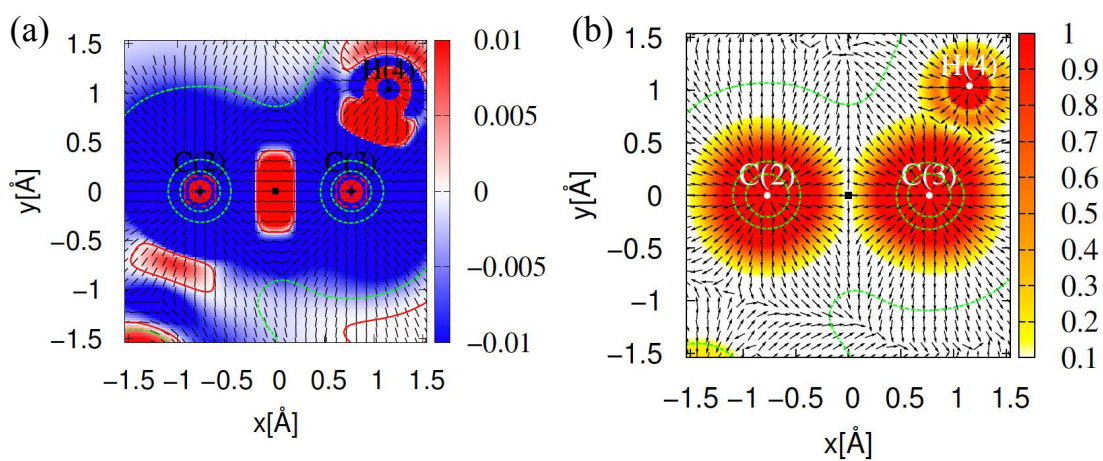}
\end{center}
\caption{The largest eigenvalue of the electronic stress tensor density (color map)
and corresponding eigenvector (black rods) are shown in panel (a), and
tension density (black arrows) and its norm (color map) are shown in panel (b), around a C-C bond in GeH$_3$(tBu). 
In both panels, the electronic interface is depicted by 
the green dashed lines, and the Lagrange point is shown by a black square. 
In panel (a), the red solid lines show the zero contour lines of the eigenvalue.}
\label{fig:GeH3tBu}
\end{figure}

\begin{figure}
\begin{center}
\includegraphics[width=14cm]{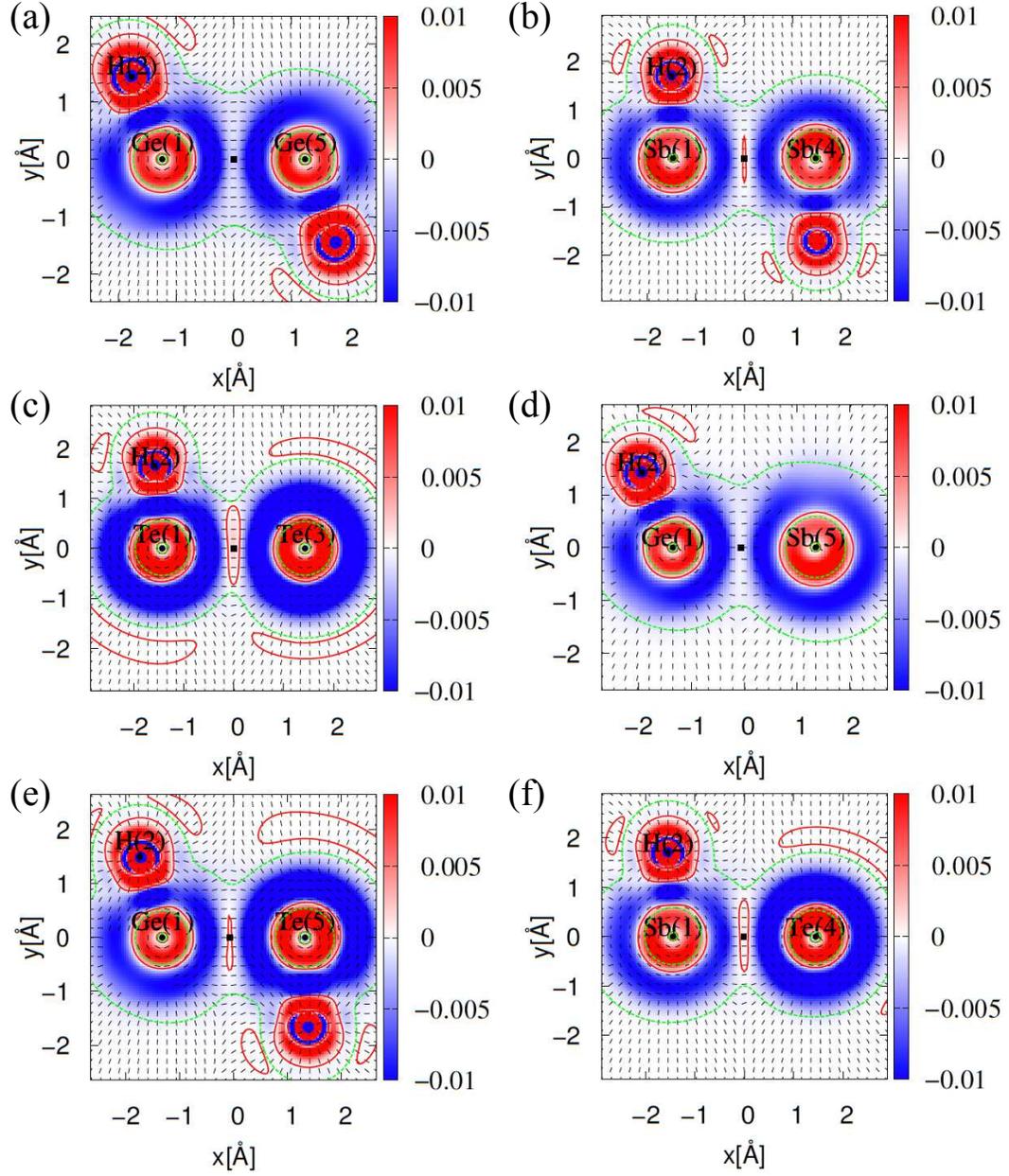}
\end{center}
\caption{The largest eigenvalue of the electronic stress tensor density and corresponding eigenvector between
the GST atoms are shown in the same manner as Fig.~\ref{fig:GeH3tBu} (a).
 (a) GeH$_3$-GeH$_3$, (b) SbH$_2$-SbH$_2$, (c) TeH-TeH, (d) GeH$_3$-SbH$_2$, (e) GeH$_3$-TeH, and 
(f) SbH$_2$-TeH.}
\label{fig:stressGST}
\end{figure}

\begin{figure}
\begin{center}
\includegraphics[width=14cm]{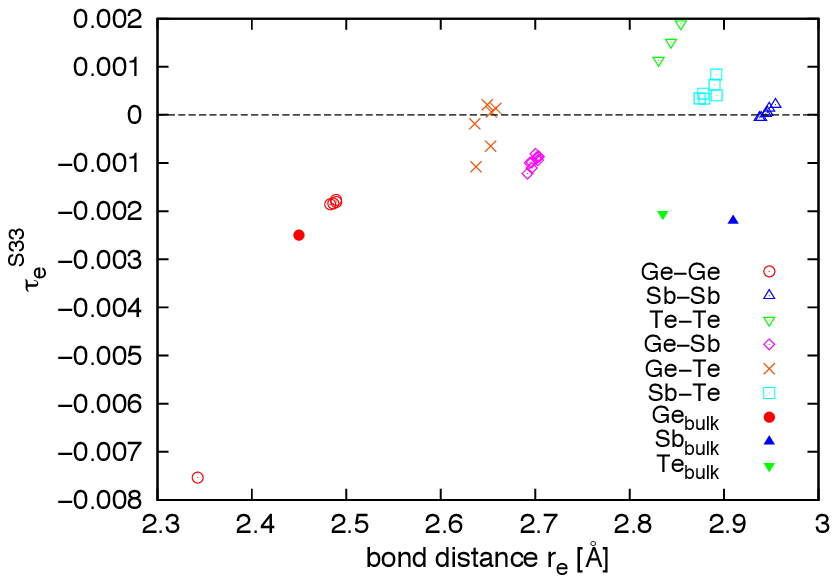}
\end{center}
\caption{The relation between bond distance and the largest eigenvalue of the electronic stress tensor at the Lagrange point.
The labels with``bulk" in the subscripts denote that they are computed for the crystal structures. }
\label{fig:dist_tau33}
\end{figure}

\begin{figure}
\begin{center}
\includegraphics[width=14cm]{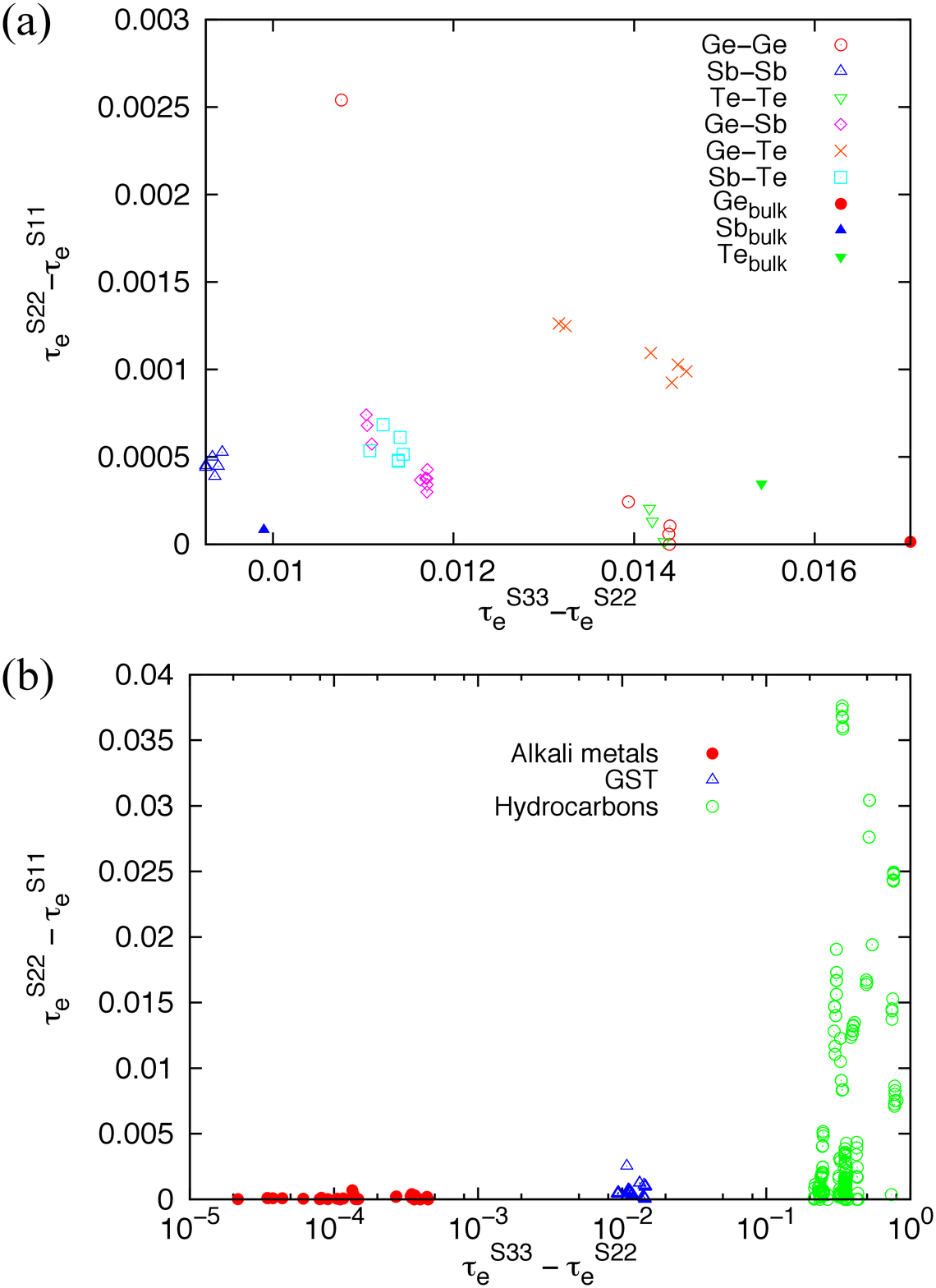}
\end{center}
\caption{Differential eigenvalues of the electronic stress tensor at the Lagrange point. In the panel (a), labels with``bulk" in the subscripts denote that they are computed for the crystal structures. }
\label{fig:diffeig}
\end{figure}

\begin{figure}
\begin{center}
\includegraphics[width=10cm]{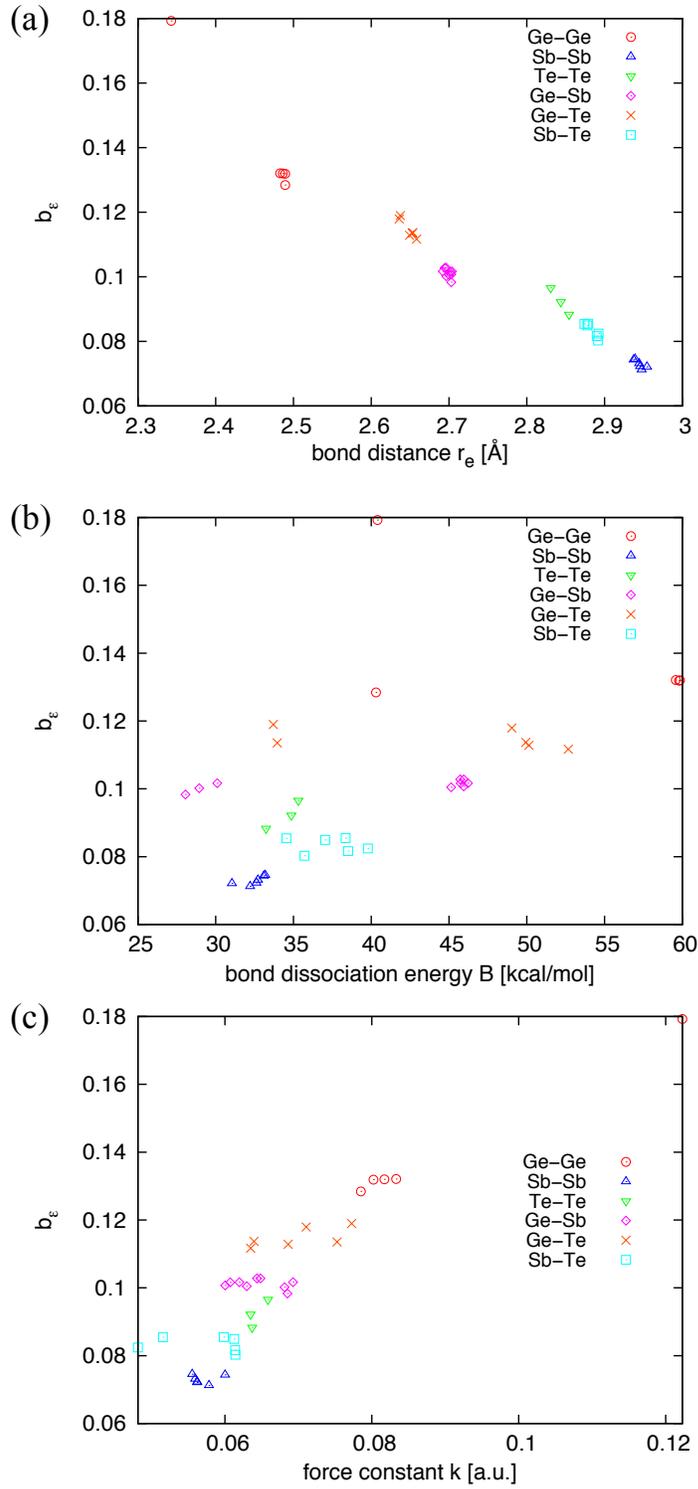}
\end{center}
\caption{The relation between energy density based bond order $b_\varepsilon$ and (a) bond distance, (b) dissociation energy, and (c) force constant, for the GST bonds.}
\label{fig:be}
\end{figure}

\begin{figure}
\begin{center}
\includegraphics[width=12cm]{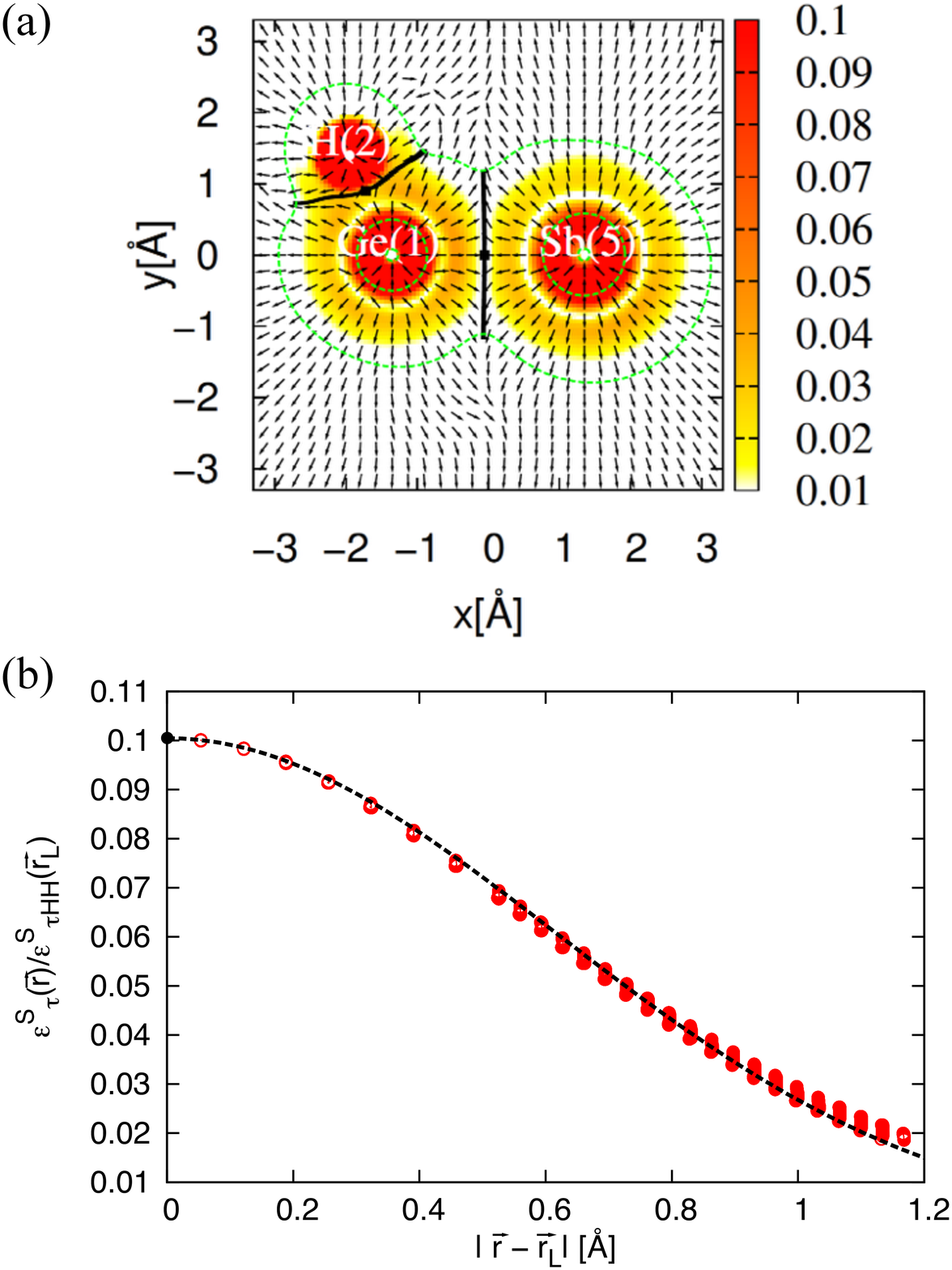}
\end{center}
\caption{In panel (a), the tension density is plotted similarly to Fig.~\ref{fig:GeH3tBu} (b) for the Ge-Sb bond in a GeH$_3$-SbH$_2$ molecule.
The Lagrange surfaces are shown by black solid lines and the Lagrange points by black squares. 
In panel (b), the energy density distribution on the Lagrange surface between the Ge and Sb atoms
is plotted as a function of the distance from the Lagrange point. The energy density is normalized by the energy density of a H$_2$ molecule at its Lagrange point. The black dashed line is the best fit Gaussian function: $b_\varepsilon \exp \left\{ -\alpha |\vec{r}-\vec{r}_L|^2 \right\}$ where $\alpha = 1.32$.}
\label{fig:GeH3SbH2}
\end{figure}

\begin{figure}
\begin{center}
\includegraphics[width=15cm]{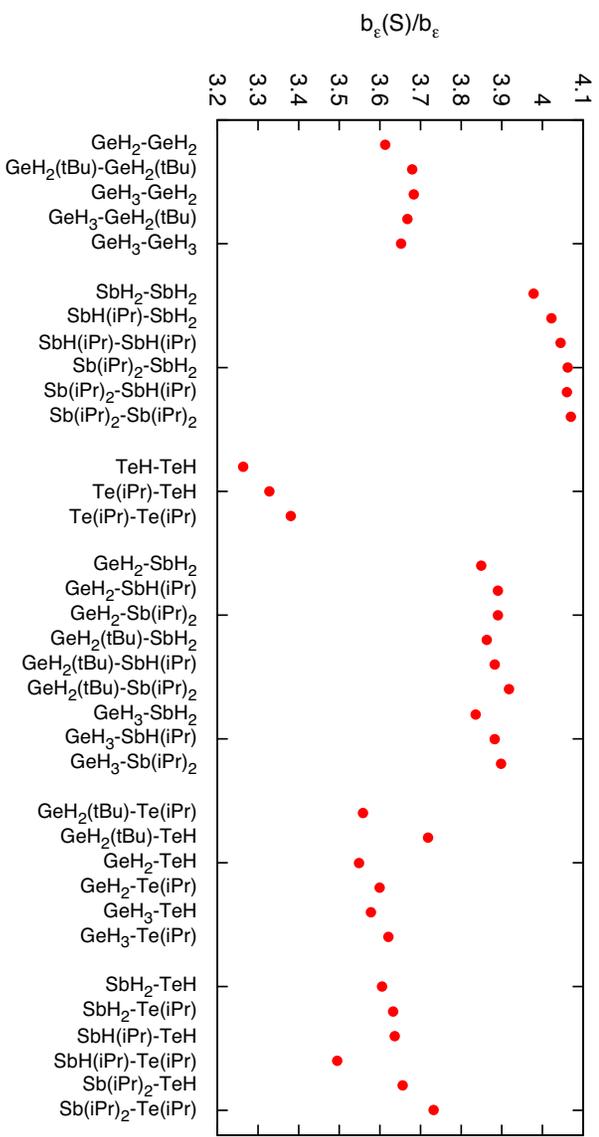}
\end{center}
\caption{The ratio of $b_{\varepsilon(S)}$ to $b_\varepsilon$ for the GST bonds.}
\label{fig:beSoverbe}
\end{figure}

\begin{figure}
\begin{center}
\includegraphics[width=15cm]{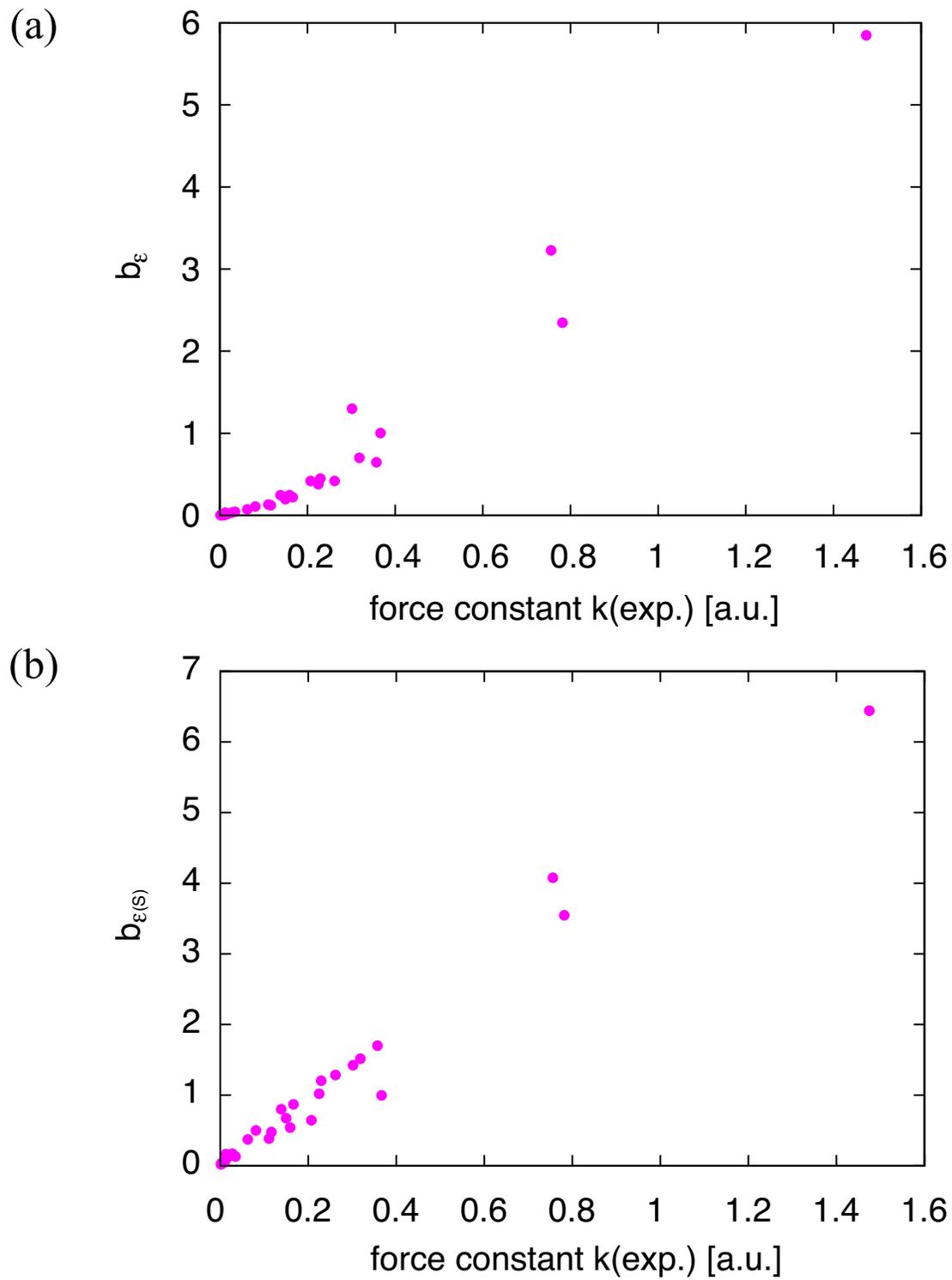}
\end{center}
\caption{The relation between force constant $k$ and (a) $b_\varepsilon$ and (b) $b_{\varepsilon(S)}$, for the bonds in the homonuclear diatomic molecules.}
\label{fig:be_DM}
\end{figure}

\fi

\clearpage

\end{document}